# Aggregated Load and Generation Equivalent Circuit Models with Semi-Empirical Data Fitting

Amritanshu Pandey[1*], Marko Jereminov[1*], Xin Li[1], Gabriela Hug[2,1], Larry Pileggi[1]

[1]Dept. of Electrical and Computer Engineering
Carnegie Mellon University
Pittsburgh, PA

[2]Power Systems Laboratory
ETH Zurich
Switzerland

*Abstract* — In this paper we propose a semi-empirical modeling framework for aggregated electrical load and generation using an equivalent circuit formulation. The proposed models are based on complex rectangular voltage and current state variables that provide a generalized form for accurately representing any transmission and distribution components. The model is based on the split equivalent circuit formulation that was previously shown to unify power flow, three phase power flow, harmonic power flow, and transient analyses. Importantly, this formulation establishes variables that are analytical and are compatible with model fitting and machine learning approaches. The parameters for the proposed semi-empirical load and generation models are synthesized from measurement data and can enable real-time simulations for time varying aggregated loads and generation.

## I. Introduction

Existing power flow and three-phase power flow algorithms predominantly use variants of constant power models (PQ and PV nodes) for aggregated load and generation modeling. The fixed power injections/absorptions (P in PV node and P, Q in PQ node) represented by these models are independent of the complex voltage at the bus. This is in stark contrast to what is observed in the field. For instance, it is shown in the B.C. Hydro system that by decreasing the substation voltage by 1%, the active and reactive power demand decreases by 1.5% and 3.4%, respectively [7]. Therefore, for a more accurate power grid analysis, a paradigm shift is needed for aggregated load and generation models to include voltage dependency. In the past, load models such as constant impedance load model, constant current load model, polynomial load model, and exponential load models [8] have been explored to address this problem. However, all of these models fall short of a generalized framework, as they are functions of complex voltage magnitude only and are independent of the complex voltage angle at the node with respect to the reference.

During normal grid operation, many nodes in the system hold the voltage magnitude constant under varying operating conditions. For instance, a generator node or a node with a reactive power-compensating device, e.g. STATCOM, static capacitors represent a typical constant voltage magnitude node to which aggregated loads and generation are connected. For these nodes, the aggregated load and generation model must also consider the voltage angle information, as the real power injection or absorption at such nodes is independent of the voltage magnitude. Therefore, the existing voltage dependent load models may fail to accurately represent aggregated load or generation connected to these nodes, as they are functions of voltage magnitude only.

Bulk installations of phasor measurement units (PMUs) on the transmission system and of remote terminal units (RTUs) on the distribution system are now providing a tremendous amount of actual electric load and generation data. In theory, this collected grid data can be used to construct accurate semi-empirical models for aggregated load and generation for power flow studies. Furthermore, these models could be automatically updated in near real-time with the latest measurement data to facilitate unprecedented simulation capabilities. Importantly, such models could assist with the compliance with FERC Order No. 693 that requires comparison of performance of the existing system in a planning power flow model to actual system behavior [6].

Toward these goals, we propose a methodology that represents the aggregated electric load and generation at a node as an equivalent circuit. This method semi-empirically formulates the generic aggregated load and generation model template as a finite order Taylor expansion of real and imaginary voltage and current state variables. Importantly, these variables correspond to the equivalent circuit elements as they are incorporated into the split-circuit formulation in [1]-[3], hence, they are analytical and suitable for application of Newton-Raphson methods.

In the proposed formulation, each circuit node is represented by expressions of rectangular complex voltage and current variables and thus, the framework can represent the aggregated load and generation as a function of both the voltage magnitude and the voltage angle with respect to the reference. Most importantly, the proposed framework models the aggregated load and generation in terms of an equivalent circuit that further allows us to apply circuit simulation methods (e.g.

SPICE [5] and its many derivatives [4]) to solve the system, such as those that are used today to simulate circuits with billions of circuit components.

Our equivalent circuit formulation for aggregated load and generation models represents a modeling template for such elements, and like other circuit abstractions, can be combined hierarchically to build larger aggregated models. Regression techniques can be used to determine the optimal coefficients for a specific instance of a template model, wherein the mean squared error is minimized between the measured currents and voltages of the system and the predicted currents and voltages of the system.

We refer to our models as Generation and Load Abstractions using Semi-empirical Surfaces, and describe the GLASS model for an induction motor (IM) and an aggregated load bus in the IEEE 390 Bus test system. Simulations are run for both the semi-empirical GLASS models and the existing physics based load models using our prototype split-circuit tool, Simulation with Unified Grid Analyses and Renewables (SUGAR). Lastly, results are compared to validate the accuracy of the developed semi-empirical models.

## II. LIMITATIONS OF EXISITING AGGREGATED LOAD AND GENERATION MODELS

At high voltage levels in the power grid, the loads must be aggregated to develop models that are suitable for computer analysis of the entire power system. Until recently, large portions of the steady-state analysis for a power grid relied on constant power load models for these aggregated representations. These constant power load models, however, have been found to be inadequate for accurate planning and security studies of the grid [8], [12]. To this effect, alternative load modeling approaches were proposed by [8] that can better capture the true behavior of the aggregated loads in the power system. These approaches model loads as a function of voltage magnitude and compute the load model parameters using measurement data from the grid. For example, the ZIP model represents the load as a mixture of constant current, constant impedance, and constant power load:

$$P_{ZIP} = P_0[a_p V^2 + b_p V + c_p] \quad (1)$$

$$Q_{ZIP} = Q_0[a_q V^2 + b_q V + c_q] \quad (2)$$

Similarly, the form of the exponential model for aggregated loads is given by:

$$P_{exp} = P_0[V^{p_v}] \quad (3)$$

$$Q_{exp} = Q_0[V^{q_v}] \quad (4)$$

Even though these load models (e.g. ZIP model, exponential model, etc.) are a significant improvement over traditional constant load models, they are still insufficient for accurately capturing certain scenarios. For example, Fig. 1 illustrates the complex voltage characteristics of a PV node in a power flow case study where the real power load connected to the PV node is varied from 100 MW to 650 MW. The graph shows that the real power absorbed by the PV node is independent of the voltage magnitude at that node and can be represented as a function of voltage angle (with respect to the reference). The load models proposed in (1) through (4) fail to capture the aggregated load characteristics for such nodes. In addition, any form of aggregated generation or distributed generation (DG) that is required to maintain a constant voltage magnitude at its node cannot be represented by existing aggregated load/generation models. Furthermore, these existing models are not compatible with transient analysis and other power systems analyses [13].

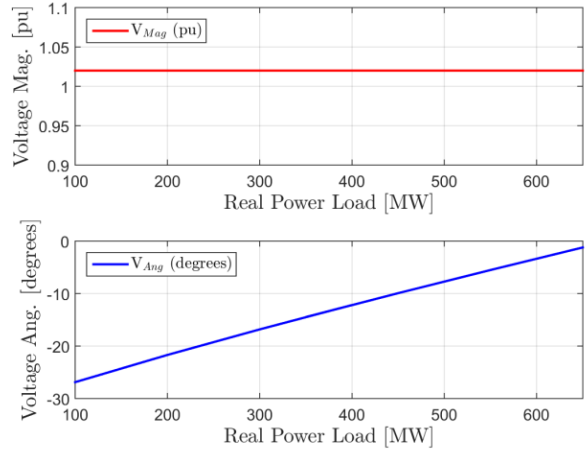

*Figure 1: Complex voltage profile on PV node with variable real power injection*

In the following sections, we will address these limitations and develop a semi-empirical aggregated load and generation model that can represent any aggregation of load and generation in the system.

## III. EQUIVALENT CIRCUIT MODEL APPROACH

The equivalent circuit approach to generalized modeling of power system's steady-state response, i.e. power flow and three-phase power flow, was recently introduced in [1]-[3]. It was shown that each of the power system components could be translated to an equivalent circuit model based on underlying relations between current and voltage state variables without loss of generality. Moreover, in order to ensure the analyticity of nonlinear complex governing circuit equations, they are split into real and imaginary parts. This splitting of complex equations corresponds to splitting of the complex equivalent circuit into two sub-circuits, i.e. a real and an imaginary, that are coupled by the controlled sources, and, hence permits the application of the Newton Raphson method to solve the circuit equations.

As an example of split-circuit modeling, consider a traditional governing equation that is used for a constant PQ load model:

$$\tilde{I}_l = \frac{P_l - jQ_l}{\tilde{V}_l^*} \quad (5)$$

The equation in (5) can be represented by a complex nonlinear voltage controlled current source, however, it contains the nonanalytic complex conjugate operator that would prevent the application of the Newton Raphson method for solving the equivalent circuit in which it was included.

Instead, we split this function into real and imaginary parts, which corresponds to an equivalent circuit model given by (6) and (7):

$$I_{Rl} = \frac{P_l V_{Rl} + Q_l V_{Il}}{V_{Rl}^2 + V_{Il}^2} \quad (6)$$

$$I_{Il} = \frac{P_l V_{Il} - Q_l V_{Rl}}{V_{Rl}^2 + V_{Il}^2} \quad (7)$$

These nonlinear split-circuit equations are further linearized via first order Taylor expansion in order to derive the complete linearized equivalent split-circuit. For instance, the Taylor expansion of the real load current in (6) about the $(i+1)^{th}$ iteration is given by:

$$I_{Rl}^{i+1} = \frac{\partial I_{Rl}}{\partial V_{Rl}}|_{V_{Rl}^i,V_{Il}^i}(V_{Rl}^{i+1}) + \frac{\partial I_{Rl}}{\partial V_{Il}}|_{V_{Rl}^i,V_{Il}^i}(V_{Il}^{i+1}) + I_{Rl}^i \\ - \frac{\partial I_{Rl}}{\partial V_{Rl}}|_{V_{Rl}^i,V_{Il}^i}(V_{Rl}^i) - \frac{\partial I_{Rl}}{\partial V_{Il}}|_{V_{Rl}^i,V_{Il}^i}(V_{Il}^i) \quad (8)$$

The first term in (8) represents a conductance, because the real load current is proportional to the real load voltage; however, the second term is proportional to the imaginary load voltage, and therefore, is represented by a voltage controlled current source. The remaining terms are all dependent on the known values from the previous iteration, so they can be lumped together and represented by an independent current source. The complete symmetric equivalent split-circuit model of the PQ load is shown in Fig. 2.

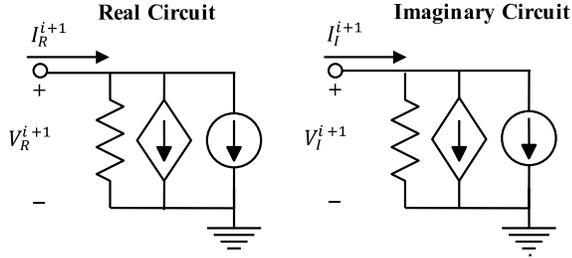

*Figure 2: Equivalent split-circuit PQ load model*

We have further shown that the previously derived split-circuit model for a PQ load [3] can be used in the three-phase power flow, where the three-phase equivalent can be obtained by connecting the per-phase equivalent circuit in Wye or Delta configuration. The three-phase grounded-Wye PQ load model is illustrated in Fig. 3 as an example.

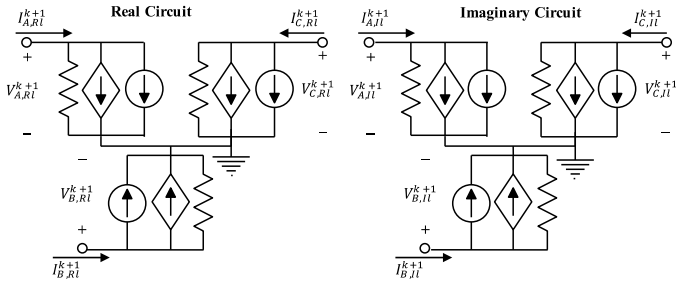

*Figure 3: Equivalent split-circuit of a three-phase grounded-Wye PQ load*

Following the same approach, each of the power system components can be represented in the equivalent split-circuit framework. Moreover, as introduced in [1]-[3], our equivalent circuit formulation based on current and voltage state variables enables the generalized modeling of any physics-based component, including complex renewable devices, as well as any practical form of nonlinear load model. This generalized modeling capability thereby enables the development of standardized accurate and robust aggregated electrical load and generation models that we further develop and characterize in the following sections.

## IV. GENERATION AND LOAD ABSTRACTIONS USING SEMI-EMPIRICAL SURFACES

Any power grid device that can be accurately characterized by current and voltage state variables can be incorporated in the equivalent circuit formulation. This formulation further enables the development of "Bottom-up" and "Top-down" semi-empirical GLASS models that can provide more accurate and robust capabilities for modeling the aggregated load and generation in the system as compared to the existing measurement based modeling approaches (PQ, ZIP, etc.) [8].

### a) "Bottom-up" GLASS model for Induction Motor

We use an induction motor (IM) to demonstrate how the semi-empirical 'bottom-up" model can be extracted from the physics based equivalent circuit of the IM and utilized for power flow and three-phase power flow analyses.

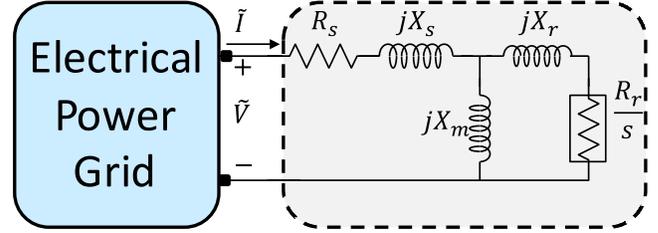

*Figure 4: Equivalent circuit representation of an IM connected to an electrical power grid.*

As shown in Fig. 4, the IM equivalent circuit as described in [10] consists of a linear RL network and a nonlinear slip ($s$) dependent resistance. The nonlinear slip dependent resistance is a function of real and imaginary components of the input voltage, $\tilde{V}$ for a given load torque. The relationship between slip and the input voltage is implicitly defined by (9) and can be explicitly obtained by solving the quadratic equation:

$$\gamma_1 s^2 + \gamma_2(V_R, V_I)s + \gamma_3 = 0 \quad (9)$$

where constants $\gamma_1, \gamma_3$ and a non-linear function of real and imaginary voltages - $\gamma_2(V_R, V_I)$ are defined as:

$$\gamma_1 = X_m^2(R_s^2 + X_s^2) \quad (10)$$

$$\gamma_2(V_R, V_I) = X_m^2 \left[2R_s R_r - \frac{3pR_r}{2\omega_s T_e}(V_R^2 + V_I^2)\right] \quad (11)$$

$$\gamma_3 = R_r^2(R_s^2 + X_s^2 + 2X_s X_m + X_m^2) \quad (12)$$

Here, $p, \omega_s, T_e$ represent the number of poles, synchronous speed in rad/s and electric torque of the induction motor in N.m, respectively.

As described by (9)-(12), the slip-dependent element from Fig. 4 can be replaced with a nonlinear voltage dependent resistance, further making the equivalent circuit dependent only on complex current and voltage state variables. Thus, the IM equivalent circuit can be reduced and represented by a nonlinear voltage dependent admittance $Y(V_R, V_I) = u(V_R, V_I) + j\,v(V_R, V_I)$ (see Fig. 5) that further defines the relationship between the input state variables as:

$$I_R + jI_I = (V_R + jV_I)[u(V_R,V_I) + jv(V_R,V_I)] \quad (13)$$

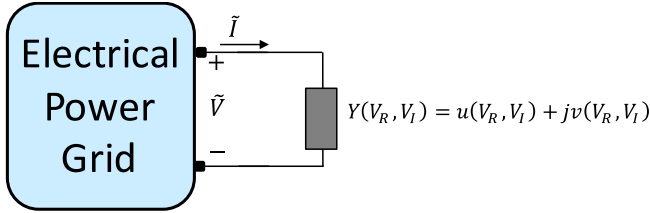

*Figure 5: Reducing the Equivalent IM circuit to obtain the relationship between input state variables*

Applying the split-circuit approach [1]-[3], the nonlinear complex current from equation (13) is split into real and imaginary currents. The equations for these currents are described via nonlinear functions of real and imaginary voltage state variables:

$$I_R = V_R u(V_R,V_I) - V_I v(V_R,V_I) \rightarrow I_R = f_R(V_R,V_I) \quad (14)$$

$$I_I = V_I u(V_R,V_I) + V_R v(V_R,V_I) \rightarrow I_I = f_I(V_R,V_I) \quad (15)$$

We further expand and approximate these nonlinear functions with a finite order Taylor expansion to obtain the two dimensional $n^{th}$ order polynomial expression given as:

$$I_C = \sum_{n=0}^{N}\left\{\frac{1}{n!}\sum_{k=0}^{n}\binom{n}{k}\nabla I_C^{n,k}(V_R - V_{R_0})^{n-k}(V_I - V_{I_0})^k\right\} \quad (16)$$

where $C$ represents the subscript placeholder for real $(R)$ or imaginary $(I)$ parts of the state variables and $N$ is the order of Taylor expansion used to accurately approximate the nonlinear function, and is dependent on desired error function. The $\nabla I_C^{n,k}$ and the binomial coefficients are defined as:

$$\nabla I_C^{n,k} = \frac{\partial^n I_C(V_R,V_I)}{\partial V_R^{n-k}\,\partial V_I^k}\bigg|_{V_{R_0},V_{I_0}} \quad (17)$$

$$\binom{n}{k} = \frac{n!}{(n-k)!\,k!} \quad (18)$$

Grouping all of the constant terms together yields the approximated expressions for both real and imaginary currents:

$$I_C = g_1^C + g_2^C V_R + g_3^C V_I + g_4^C V_R V_I + g_5^C V_R^2 + g_6^C V_I^2 \cdots \quad (19)$$

where the optimal values for $g_i^C$ coefficients will be determined by the standard least-squares fitting techniques [9].

*b) "Top-down" Generic GLASS model for Aggregated Load and Generation*

Without loss of generality, we base "Top-down" GLASS models on the assumption that the steady-state response of every aggregated load and generation abstraction can be represented by functions of current and voltage state variables, and fit the model using finite order Taylor expansion. The governing equation of the model can be obtained from the generalized two-dimensional Taylor expansion:

$$A_C = \sum_{n=0}^{N}\left\{\frac{1}{n!}\sum_{k=0}^{n}\binom{n}{k}\nabla A_C^{nk}(B_R - B_{R_0})^{n-k}(B_I - B_{I_0})^k\right\} \quad (20)$$

where $\nabla A_C^{nk}$ is given as:

$$\nabla A_C^{nk} = \frac{\partial^n A_C(B_R,B_I)}{\partial B_R^{n-k}\,\partial B_I^k}\bigg|_{B_{R_0},B_{I_0}} \quad (21)$$

The GLASS model captures load or generation as a function of voltage or current. By choosing the $\{A_R, A_I\}$ as a set of current state variables, real and imaginary voltages are defined as dependent state variables $\{B_R, B_I\}$ and by choosing the $\{A_R, A_I\}$ as a set of voltage state variables, real and imaginary currents are defined as dependent state variables $\{B_R, B_I\}$.

Following the same methodology as in the derivation of the induction motor model, the expansion from (20) can be written in the following compressed form:

$$A_C = g_1^C + g_2^C B_R + g_3^C B_I + g_4^C B_R B_I + g_5^C B_R^2 + g_6^C B_I^2 \cdots \quad (22)$$

To obtain the generalized equivalent split circuit model we further linearize the nonlinear functions in (22) around the values obtained in the $(i+1)^{th}$ iteration:

$$A_C^{i+1} = \frac{\partial A_C}{\partial B_R}\bigg|_{B_R^i,B_I^i}(B_R^{i+1}) + \frac{\partial A_C}{\partial B_I}\bigg|_{B_R^i,B_I^i}(B_I^{i+1}) + A_C^i$$
$$- \frac{\partial A_C}{\partial B_R}\bigg|_{B_R^i,B_I^i}(B_R^i) + \frac{\partial A_C}{\partial B_I}\bigg|_{B_R^i,B_I^i}(B_I^i) \quad (23)$$

Defining the real and imaginary voltages as dependent variables, i.e. $\{B_R, B_I\} \rightarrow \{V_R, V_I\}, \{A_R, A_I\} \rightarrow \{I_R, I_I\}$, the expressions in (23) can be represented by the equivalent split circuit shown in Fig. 6. From the figure, the terms dependent on the voltage across the respective circuit, i.e. $\frac{\partial I_C}{\partial V_R}\big|_{V_R^i,V_I^i}(V_R^{i+1}), \frac{\partial I_C}{\partial V_I}\big|_{V_R^i,V_I^i}(V_I^{i+1})$ and the terms dependent on the voltage across the other circuit, i.e. $\frac{\partial I_C}{\partial V_R}\big|_{V_R^i,V_I^i}(V_I^{i+1})$, $\frac{\partial I_C}{\partial V_I}\big|_{V_R^i,V_I^i}(V_R^{i+1})$ are represented by a conductance and a voltage controlled current source respectively. Furthermore, the terms dependent on the values from the previous iteration are lumped together and are represented by an independent current source. Similarly, defining the currents as dependent variables can be further translated into the equivalent circuit presented in Fig. 7. In this figure, the terms dependent on the current through itself, i.e. $\frac{\partial V_C}{\partial I_R}\big|_{I_R^i,I_I^i}(I_R^{i+1}), \frac{\partial V_C}{\partial I_I}\big|_{I_R^i,I_I^i}(I_I^{i+1})$ and the terms dependent on the current in the other circuit, i.e. $\frac{\partial V_C}{\partial I_R}\big|_{I_R^i,I_I^i}(I_I^{i+1}), \frac{\partial V_C}{\partial I_I}\big|_{I_R^i,I_I^i}(I_R^{i+1})$ are represented by a resistor and a current controlled voltage source respectively, while the terms dependent on the values from the previous iteration are lumped together and replaced by an independent voltage source.

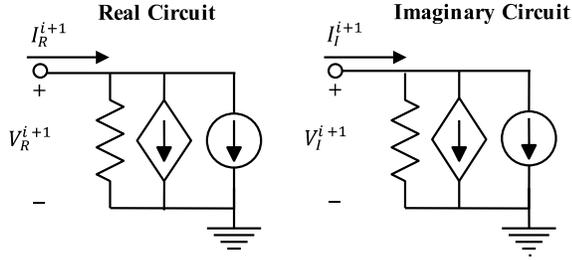

*Figure 6: GLASS equivalent split circuit for voltage dependent variables*

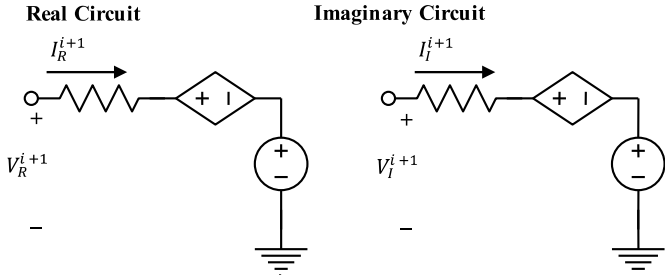

*Figure 7: GLASS equivalent split circuit for current dependent variables*

## V. RESULTS AND DISCUSSION

As a proof of concept, the proposed semi-empirical modeling formulation of aggregated load and generation in the power grid is validated with two study cases: i) steady-state analysis of an IM using the "Bottom-up" GLASS model; ii) steady-state analysis of the IEEE 390 test system using the "Top-down" generic GLASS model.

In the first study case, the steady-state behavior of the IM is simulated by a GLASS model that is based on the third order Taylor expansion. In order to compute the GLASS model coefficients, synthetic measurement data was first created by running time-domain simulations on the physics based model of IM in SUGAR. The generated synthetic measurement data was then used as an input into the MATLAB optimization toolbox [14],[15] to generate the model coefficients. Some data examples of the GLASS model's coefficients computed from synthetic measurement data are shown in Table 1. Once the semi-empirical GLASS model for IM was developed, simulation was run on it for different source voltages and load torques and the results from the simulation were compared against the synthetic measurement data from time-domain simulations in SUGAR.

TABLE 1: IM GLASS TEMPLATE COEFFICIENTS FOR LOW TORQUE AND HIGH TORQUE MODEL

| Model | | $g_1^C$ | $g_2^C$ | $g_3^C$ | $g_4^C$ | $g_5^C$ | $g_6^C$ | $g_7^C$ | $g_8^C$ |
|---|---|---|---|---|---|---|---|---|---|
| Low Torque | $I_R$ | 63.483 | -0.345 | # | # | 0.001 | # | -7.09E-07 | # |
| | $I_I$ | -30.235 | 0.212 | # | # | -0.001 | # | 6.72E-07 | # |
| High Torque | $I_R$ | 106.950 | -0.591 | # | # | 0.001 | # | -1.21E-06 | # |
| | $I_I$ | -69.319 | 0.486 | # | # | -0.001 | # | 1.19E-06 | # |

**Note**: # represents coefficients that are inconsequential for this simulation as a result of measured $V_I$ being of zero magnitude

Comparison between the fundamental frequency component of the source voltage and load current for the GLASS model simulation and the synthetic measurement data is presented in Fig. 8. The high accuracy of the GLASS model is validated as the simulation results from the GLASS model closely match the synthetic measurement data. Furthermore, the graph in Fig. 8 shows the ability of the GLASS model to capture the true system conditions under external disturbances. For example, the GLASS model predicts the correct load currents of IM for a torque change at t=5 sec when the load torque of the motor is doubled. Table 2 documents the real and imaginary voltages and currents that were used to plot the graphs in Fig. 8. The data shows the accuracy to within three significant digits. Importantly, this comparison demonstrates the high accuracy of the GLASS model for a range of voltages that were initially not used to create the model.

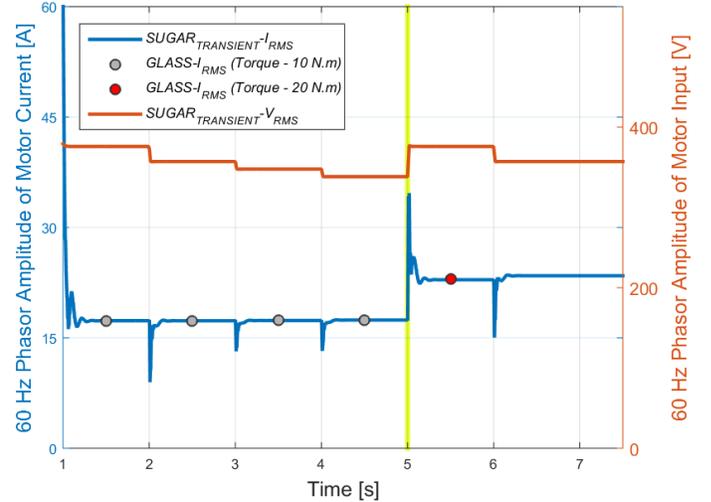

*Figure 8: Comparison of GLASS results for the IM against synthetic measurement data created using SUGAR's IM model*

TABLE 2: COMPARISON OF RESULTS BETWEEN THE GLASS LOAD MODEL FOR THE IM AND THE SYNTHETIC MEASUREMENT DATA OF THE IM

| Model Type | Real Voltage ($V_R$) | Imag. Voltage ($V_I$) | Real Current ($I_R$) | Imag. Current ($I_I$) | Load Torque ($T_L$) |
|---|---|---|---|---|---|
| GLASS | **375.59** | **0.00** | **11.31** | **-13.03** | **10** |
| Syn. Meas. | 375.59 | 0.00 | 11.36 | -13.09 | 10 |
| GLASS | 356.81 | 0.00 | 11.95 | -12.54 | 10 |
| Syn. Meas. | 356.81 | 0.00 | 11.95 | -12.54 | 10 |
| GLASS | **347.42** | **0.00** | **12.28** | **-12.28** | **10** |
| Syn. Meas. | 347.42 | 0.00 | 12.27 | -12.27 | 10 |
| GLASS | 338.03 | 0.00 | 12.62 | -12.02 | 10 |
| Syn. Meas. | 338.03 | 0.00 | 12.61 | -12.01 | 10 |
| GLASS | **375.59** | **0.00** | **18.27** | **-13.89** | **20** |
| Syn. Meas. | 375.59 | 0.00 | 18.20 | -13.89 | 20 |

**Note:** The numbers in bold represent the extrapolated results of the GLASS load model for the IM

For the second case study, the individual loads in the IEEE 390 Bus test system as seen from the source node are aggregated using the generic GLASS model. Time-domain simulations were first performed on the IEEE 390 Bus test system in EMTP-RV [11] to generate synthetic measurement data. The synthetic measurement data was then used to compute the generic GLASS model coefficients using the MATLAB optimization toolbox. Once the coefficients for the

GLASS model were computed for the IEEE 390 bus aggregated loads, a simulation was run to simulate a sudden voltage drop in the system due to a disturbance on the source node. The plots in Fig. 9 present the simulation results that compare the voltages and currents between the synthetic measurement data collected from the source node in EMTP-RV simulation and those produced by the GLASS model for the aggregated load. The results from the plot show that the steady-state load currents produced by GLASS model closely match the synthetically measured load current in the EMTP-RV simulation environment.

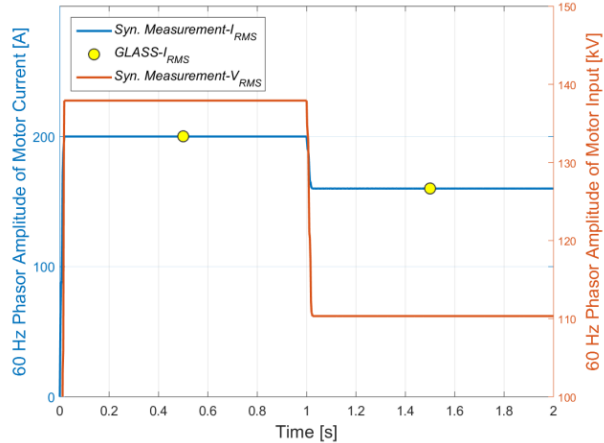

*Figure 9: Comparison of GLASS results for the IEEE 390 Bus Test Case with synthetic measurement data created using EMTP-RV*

Table 3: GLASS COEFFICIENTS FOR IEEE 390 BUS TEST CASE

| Complex Var. | $g_1^C$ | $g_2^C$ | $g_3^C$ |
|---|---|---|---|
| $I_R$ | 0.0932 | -8.86E-04 | 0.0014 |
| $I_I$ | -0.170 | -0.0012 | -0.0035 |

## VI. CONCLUSION AND FUTURE WORK

In this paper, we developed aggregated load and generation models for power system analyses based on a semi-empirical modeling approach. It was shown that finite order Taylor expansion with the use of complex rectangular voltage and current state variables could model any power grid component that is described by such variables.

As future work, we propose to extend our aggregated load and generation modeling framework to model time dependencies for aggregated loads and frequency dependent aggregated non-linear loads for harmonic balance analysis. We further plan to explore better data fitting capabilities based on the latest machine learning methods that could enable models that include speculative variables. The ultimate goal would be models that enable automated real-time simulation wherein measurement-based load and generation models can be updated in near real-time to assist with system operation, reliability, and security.

ACKNOWLEDGMENT

This research is funded in part by the CMU-SYSU Collaborative Innovation Research Center and the SYSU-CMU International Joint Research Institute.